\documentclass[10pt, amssymb,preprint,twocolumn]{revtex4}
\usepackage{hyperref}
\usepackage{amsmath}
\usepackage{graphicx}
\usepackage{amssymb} 
\usepackage{amsfonts}
\begin{document}

\title{Overview of the structural unification of quantum mechanics and relativity using the algebra of quantions}
\author{Florin Moldoveanu}
\affiliation{Department of Theoretical Physics, National Institute for Physics and Nuclear Engineering, PO Box MG-6, Bucharest, Romania $^{\dag}$\footnotetext{$\dag$ On leave.}}
\email{fmoldove@gmail.com}

\begin{abstract}
The purpose of this contribution is to provide an introduction for a general physics audience to the recent results of Emile Grgin that unifies quantum mechanics and relativity into the same mathematical structure. This structure is the algebra of quantions, a non-division algebra that is the natural framework for electroweak theory on curved space-time. Similar with quaternions, quantions preserve the core features of associativity and complex conjugation while giving up the unnecessarily historically biased property of division. Lack of division makes possible structural unification with relativity (one cannot upgrade the linear Minkowski space to a division algebra due to null light-cone vectors) and demands an adjustment from Born's standard interpretation of the wave function in terms of probability currents. This paper is an overview to the theory of quantions, followed by discussions.\end{abstract}


\maketitle
\section{Introduction} Unification of quantum mechanics and general relativity is the main challenge of today's physics. Several approaches have been proposed so far with various degrees of success: string theory \cite{stringTheory}, loop quantum gravity \cite{LoopQGrav}, twistor theory \cite{twistor}, and non-commuting geometry \cite{Connes}. The root cause of the tension between quantum mechanics and relativity stems from the difference in the underlying Lie groups and Lie algebras: unitary groups for quantum mechanics, and orthogonal groups for relativity.  This paper will present some of the core results of a relatively new approach towards unification pioneered by Emile Grgin: structural unification of quantum mechanics and relativity based on the algebra of quantions \cite{GrginBook1}. This is an overview of those results aimed at presenting the material for a general physics audience. 

There are several points of view that can illustrate quantions. We will start with the historical account and original justification for quantions. Basic algebraic properties will be presented. Then quantions can be described mathematically as the algebra that removes a degeneracy of the complex numbers. Next, Born's interpretation of the wave function admits an interesting geometric interpretation, and deformations of the geometry were considered in the past as a way to search for relativity and quantum mechanics unification. In quantionic physics, Born's interpretation is naturally generalized and replaced by Zovko's interpretation which leads directly to Dirac's equation and the semi-classical aspect of the electroweak theory$^{1}$\footnotetext[1]{Second quantization is not yet researched in this approach.}. Because electroweak physics follows as a theorem from quantionic properties, quantions are a major step towards the axiomatization of physics. Last, the open problems are considered in an extended discussion section. The author's conjectures about quantions and a possible new physics paradigm are presented as well.

\section{Quantions, the historical perspective} Structural unification of quantum mechanics and relativity started with a collaboration between Emile Grgin and Aage Petersen and was rooted into Bohr's belief that the correspondence principle has more secrets to reveal. Acting on this belief, Bohr's assistant Aage Petersen in collaboration with Emile Grgin started looking at the common elements of classical and quantum mechanics. The idea was that classical and quantum mechanics shared characteristics reveal core physics features that are otherwise obscured by the (non-essential) details related to the realization of those theories in phase and Hilbert space respectively. The resulting mathematical structure called a ``quantal algebra'' is a unification of a Poisson algebra with a Lie-Jordan algebra, a result also obtained by other authors \cite{LieJordan}. 

Quantal algebra is rooted into two postulates, or observations which can be made about classical and quantum mechanics. The first observation is that classical and quantum mechanics use two products: one symmetric and one anti-symmetric. For example, in the classical case one has the regular multiplication and the Poisson bracket. In the usual formulation of quantum mechanics, one has the anti-commutator (the Jordan product) and the commutator. In phase space, quantum mechanics is described by the cosine and sine Moyal brackets \cite{GrginBook1}. The second observation was that classical and quantum mechanics obey the so-called composability principle: any two physical systems can interact with each other. When two physical systems interact we need to preserve the original structure, meaning the symmetric and anti-symmetric products. Let us call $S_1$ and $A_1$ the symmetric and anti-symmetric products of system $1$, $S_2$ and $A_2$ the corresponding products of system $2$, and $S_T$ and $A_T$ the products of the total system. Then composability implies:
\begin{eqnarray}
S_T = S_1 S_2 - a A_1 A_2    
\label{S1}\\
A_T = A_1 S_2 + S_1 A_2
\label{A1}
\end{eqnarray}

where $a$ could be $1$, $0$, or $-1$ \cite{composabilityPaper}.

Comparing Eqs.~\ref{S1}  and \ref{A1} with complex number multiplication, it is easy see that when $a = 1$ one can identify $S$ with the real part, and $A$ with the imaginary part of a complex number. Detail analysis reveals that $a = \hbar^2$ for quantum mechanics and $a = 0$ for classical mechanics. The $a = -1$ case would correspond to a quantum mechanics based on split-complex numbers. This case might be considered unphysical because split complex numbers (which use $j^2 = 1$) do not satisfy the spectral theorem which gives uniqueness to quantum mechanics \cite{vonNewman}. In general, in quantum mechanics $S$ is a product in the space of observables $\cal{O}$ and $A$ is a product in the space of abstract generators $\cal{L}$. Hermitian matrices represent observables, while anti-hermitian matrices represent generators. Since Hermitian and anti-hermitian matrices are in one-to-one correspondence, it is tempting to postulate the equivalence of $\cal{O}$ and $\cal{L}$, but in fact this is just a natural consequence of the composability principle. In terms of interpretation of quantum mechanics, the origin of complex numbers is a very unintuitive feature of quantum mechanics. From Eqs.~\ref{S1} and \ref{A1} it is easy to understand it as a structure preserving requirement under composability. Also, the naive limit $\hbar \rightarrow 0$ that is typically assumed to describing the transition from quantum to classical mechanics is replaced with the correct exact structural transition $\hbar ^2 = 0$ to a nilpotent algebra.

Combining classical and quantum mechanics into a unified structure called a quantal algebra (a term coined by Peterson and Grgin), and renaming the symmetric and the anti-symmetric product as $\sigma$ and $\alpha$ respectively, one has the following requirements:
\begin{eqnarray}
(f \alpha g) \alpha h + (g\alpha h) \alpha f + (h \alpha f) \alpha g = 0
\label{JacobiIdentity}\\
g \alpha (f \sigma h) = (g \alpha f) \sigma h + f \sigma (g \alpha h)
\label{LeibnitzIdentity}\\
(f \sigma g) \sigma h - f \sigma (g \sigma h) = a g \alpha (h \alpha f)
\label{PetersenIdentity}
\end{eqnarray}	

The difference between a quantal algebra and a Lie-Jordan algebra is that a Lie-Jordan algebra has additional properties relating to its spectral properties\cite{LieJordan}. Those properties eliminate the need for the split-complex numbers and they are not derived from the composability principle. In the following, unless we explicitly specify it, we will restrict the discussion to only the quantum mechanics case of $a = 1$.

Eq.~\ref{JacobiIdentity} represents the usual Jacobi  identity and captures the Lie part of the quantal algebra. Eq.~\ref{LeibnitzIdentity} is the distribution law of the Lie over the Jordan product and can be understood in terms of infinitesimal automorphisms. Suppose that $T = I + \epsilon F \alpha$ is an infinitesimal automorphism. Then infinitesimal motions in the quantal algebra must be compatible with the algebraic product sigma:
$T(f \sigma g) = (Tf)\sigma(Tg)$. This simplifies to the Leibniz identity: 
$F \alpha (f \sigma g) = (F \alpha f) \sigma g + f \sigma (F \alpha g)$

In general, a Jordan algebra is non-associative. Introducing the associator as a measure of non-associativity:
\begin{equation}
[f,g,h] = (f \sigma g) \sigma h - f \sigma (g \sigma h)
\label{assocdef}
\end{equation}
then Eq.~\ref{PetersenIdentity}, (proposed to be called ``the Petersen's identity" by Emile Grgin), can be written as \cite{GrginBook1}:
\begin{equation}
[f,g,h]_{\sigma} = a g \alpha (h \alpha f)
\label{Petersen2}
\end{equation}

In general, one can construct a mapping $J$ between $\cal{O}$ (less the unit element) and $\cal{L}$:
\begin{equation}
J: \cal{O} \rightarrow \cal{L}
\end{equation}
such that:
\begin{equation}
F = J f 
\end{equation}
where $f  \in \cal{O}$ and $F \in \cal{L}$, and:
\begin{equation}
f = - a J F
\end{equation}
which for quantum mechanics implies:
\begin{equation}
J J = -I
\end{equation}

If one introduces a new product beta defined as:
\begin{equation}
f \beta g = f \sigma g + i f \alpha g
\label{externalComplexification}
\end{equation}
then $\beta$ is an associative product. There are two ways to introduce the associative product. The typical way, (called external complexification by Grgin), follows the prescription of Eq~\ref{externalComplexification}. However, there is another way, (called internal complexification$^{2}$\footnotetext[2]{Because internal complexification is the critical idea of quantionic research, I propose to call it the Grgin complexification.} by Grgin \cite{GrginBook1}). In this case one element of the algebra will play the role of  $\sqrt{-1}$. Let us assume that $J = \sqrt{-e}$ in $\cal{O}$ exists. If $\cal{O}_J$ is the centralizer of $J$, i.e. the set of all observables $f$ in $\cal{O}$ such that $J \alpha f = 0$, then $\{\cal{O}_J, \sigma , \alpha , \it{e} \}$ is a quantal algebra. $J$ may, or may not exist, but if it does, $J$ plays a unique role in the algebra, and will later introduce relativity into the quantal framework. From Eq.~\ref{S1} it is easy to see that the spectral characteristics are defined only by the symmetric product (due to the choice of complex, or split complex numbers based on $a$). Quantum mechanics and relativity share Jordan algebra characteristics \cite{BertramPaper}. 

At this point, it is useful to review Lie algebras \cite{JohnBaez}and Lie groups. Lie groups are manifolds endowed with group properties. Lie algebras are associated with the tangent space of the Lie group at the identity element. Different Lie groups can share the same Lie algebra, and there are Lie algebras which do not correspond to any Lie group. There are four infinite families of ``classical" simple Lie algebras: unitary algebras $su(n+1)$ (A series), odd orthogonal algebras $so(2n+1)$ (B series), symplectic algebras $sp(2n)$ (C series), and even orthogonal algebras $so(2n)$ (D series). In addition to those, there are five ``exceptional" simple Lie algebras: $g_2$, $f_4$, $e_6$, $e_7$, and $e_8$. In terms of normed division number systems over the real numbers, the orthogonal algebras correspond to real numbers $\mathbb R$, the unitary algebras correspond to complex numbers $\mathbb C$, the symplectic algebras correspond to quaternions $\mathbb H$, and the exceptional algebras correspond to the non-associative octonions $\mathbb O$, terminating the series.

One way to analyze Lie algebras is by the Cartan classification based on the Jacoby identity (Eq.~\ref{JacobiIdentity}). When one imposes the additional constraints of Eqs.~\ref{LeibnitzIdentity} and \ref{PetersenIdentity}, then one expects a restriction in terms of possible Lie algebras. Grgin identified four cases: the infinite family of unitary algebras $su(n+1)$ and three ``sporadic'' orthogonal algebras: $so(3)$, $so(6)$, and $so(2,4)$ \cite{foundationPaper1,  foundationPaper2, foundationPaper3, foundationPaper4}. Since the Lie group $SO(3)$ is isomorphic with the $SU(2)$ group and $SO(6)$ is isomorphic with $SU(4)$, the only case that does not reduce itself to standard non-relativistic quantum mechanics is $so(2,4)$. The Lie group $SO(2,4)$ corresponds to the conformal compactification of the Minkowski space, is isomorphic with $SU(2,2)$, and leads to Penrose's twistor theory \cite{twistor}. The Lie algebra $so(2,4)$ leads to the algebra of quantions and is the unique mathematical structure that contains both quantum mechanics (a quantal algebra) and relativity in exactly four dimensions. Since the translation group does not appear in quantionic algebra, the space is intrinsic Riemannian, and quantionic physics structurally unifies quantum mechanics with general relativity. Wolfgang Bertram identified another family of quantal algebra realization, the pseudo-unitary $u(p,q)$ algebras of indefinite signature \cite{BertramCommunication}. He also pointed out that quantal algebras are basically $C^*$ algebras with no positivity condition.

But what is the heuristic reason for using internal complexification in the first place, and why does it lead to relativity? As seen from Cartan's classification, we have only symplectic, unitary, and orthogonal algebras. A quantal algebra contains the symplectic and unitary ingredients by default because it unifies classical and quantum mechanics. Relativity requires orthogonal algebras and non divisibility. If we can obtain a generalization of complex numbers that is not isomorphic with a unitary group (which implies divisibility), then it must contain some form of orthogonal algebra with the hope that maybe relativity will somehow arise from it. For a Hermitian matrix $H$, one has: $Tr(H^2) > 0$ and  $Tr(-I) < 0$ and therefore standard complexification does not contain generalizations of complex numbers. Only internal complexification can lead to non-unitary quantal algebras and $so(2,4)$ is the only possible orthogonal solution. To obtain relativity, recall that we are looking only at a subset space defined by the constraint: $J \alpha f = 0$. Once $J$ is selected, quantions are defined into a subspace of $so(2,4)$, the centralizer space $O_J (2,4)$. The centralizer reduces itself to a complex Minkowski space of dimensionality $8$: $M_0 (\mathbb{C}) = M_0 \oplus i M_0$ and any element $f\in O_J (2,4)$ is of the form:
\begin{equation}
f = f_r + J \beta f_i
\label{newNumberSystem}
\end{equation}
with $f_r$ and $f_i$ real. 
The linear space $L^{(2,4)}$ on which the group $SO(2,4)$ acts, is a distinguished unique space, because only in this case one can define uniquely complex conjugation as a reflection that cannot be undone by continuous transformations. 
	
\subsection{Algebraic properties of quantions}
Let us explore same basic properties of the quantions. This section will follow closely the quantionic book of Emile Grgin \cite{GrginBook1}.  The first observation is that $J = \sqrt{(-e)}$ is not unique. There are an infinity of solutions of dimensionality 3 which are transitively related by the $SO(1,3)$ group. The algebraic unit $e$ of quantion algebra $\mathbb D$ is a contravariant complex four vector that defines the time direction in the local frame. 

In terms of complex numbers, a quantion is a $2\times 2$ matrix
\begin{math}
\left(
\begin{tabular}{cc}
z & v\\
u & w\\
\end{tabular}
\right)
\end{math}
with the following multiplication rule:
\begin{equation}
\left(
\begin{tabular}{cc}
a & c\\
b & d\\
\end{tabular}
\right)
*
\left(
\begin{tabular}{cc}
z & v\\
u & w\\
\end{tabular}
\right)
=
\left(
\begin{tabular}{cc}
az + cu & av + cw\\
bz + du & bv + dw\\
\end{tabular}
\right)
\end{equation}

Using the Minkowski scalar product:
\begin{equation}
(u, v) \equiv \eta_{\mu \nu} u^{\mu}v^{\nu}
\end{equation}
where $\eta_{\mu \nu} = diag (1, -1, -1, -1)$
and renaming the unit $e$ as $\Omega$, the product $\beta$ is:
\begin{equation}
u \beta v = (\Omega , u) v + (\Omega , v) u - (u, v) - i * (\Omega \wedge u \wedge v)
\end{equation}
where $*$ is the Hodge duality mapping.

In general, one can decompose any arbitrary quantion in the following form:
\begin{equation}
u = U \Omega + \overrightarrow{u}
\end{equation}

If we introduce $\Pi$ as the 3-dimensional hyperplane orthogonal to $\Omega$, and choosing a set $\{\overrightarrow{e_1}, \overrightarrow{e_2}, \overrightarrow{e_3}\}$ of orthonormal vectors in $\Pi$, then the multiplication table for $\beta$ is:
\begin{equation}
\begin{tabular}{|c|c|c|c|c|c|}
\hline
$~~\beta~~$ && $~~\Omega~~$ & $\overrightarrow{e_1}$ & $\overrightarrow{e_2}$ & $\overrightarrow{e_3}$ \\
\hline
\hline
$\Omega$ && $\Omega$ & $\overrightarrow{e_1}$ & $\overrightarrow{e_2}$ & $\overrightarrow{e_3}$ \\
\hline
$\overrightarrow{e_1}$ && $\overrightarrow{e_1}$ & $\Omega$ & $i \overrightarrow{e_3}$ & $-i \overrightarrow{e_2}$ \\
\hline
$\overrightarrow{e_2}$ && $\overrightarrow{e_2}$ & $-i \overrightarrow{e_3}$ & $\Omega$ & $i \overrightarrow{e_1}$ \\
\hline
$\overrightarrow{e_3}$ && $\overrightarrow{e_3}$ & $i \overrightarrow{e_2}$ & $-i \overrightarrow{e_1}$ & $\Omega$ \\
\hline
\end{tabular}
\end{equation}

This multiplication table is identical with the Pauli matrices multiplication table with the following identification: $( \Omega \leftrightarrow \sigma_0, \overrightarrow{e_i} \leftrightarrow \sigma_i )$. Hence, in a fixed tetrad, the algebra of quantions can be represented by the algebra of $2 \times 2$ complex matrices. This is because the Lorenz group is isomorphic with $SL(2, \mathbb{C})$. Expressed in terms of Pauli matrices, a quantion can be written as:
\begin{equation}
q = q_0 I +  \overrightarrow{q} . \overrightarrow{\sigma}
\end{equation}
This form was first studied by James Edmonds \cite{EdmondsPaper} in 1972. 

Quaternionic multiplication table is:
\begin{equation}
\begin{tabular}{|c|c|c|c|c|c|}
\hline
$~~.~~$ && $~~1~~$ & $\overrightarrow{i}$ & $\overrightarrow{j}$ & $\overrightarrow{k}$ \\
\hline
\hline
$1$ && $1$ & $\overrightarrow{i}$ & $\overrightarrow{j}$ & $\overrightarrow{k}$ \\
\hline
$\overrightarrow{i}$ && $\overrightarrow{i}$ & $-1$ & $\overrightarrow{k}$ & $- \overrightarrow{j}$ \\
\hline
$\overrightarrow{j}$ && $\overrightarrow{j}$ & $- \overrightarrow{k}$ & $-1$ & $ \overrightarrow{i}$ \\
\hline
$\overrightarrow{k}$ && $\overrightarrow{k}$ & $\overrightarrow{j}$ & $- \overrightarrow{i}$ & $-1$ \\
\hline
\end{tabular}
\end{equation}

Comparing quaternions to quantions, the transformation rule between the two algebras is:
\begin{equation}
\begin{array}{rcl}
\Omega &=&1\\
i \overrightarrow{e_1}&=& \overrightarrow{i}\\
i \overrightarrow{e_2}&=& \overrightarrow{j}\\
i \overrightarrow{e_3}&=& \overrightarrow{k}\\
\end{array}
\end{equation}

The linear spaces of real quantions and real quaternions are different four-dimensional slices of the algebra of complex quaternions.

Given the tetrad $\{ \Omega, \overrightarrow{e_1}, \overrightarrow{e_2}, \overrightarrow{e_3}\}$, let us introduce the null tetrad $\{ l, n, m, \overline{m} \}$ by the relations:
\begin{equation}
\begin{array}{rcl}
l &=&\frac{1}{2} (\Omega + \overrightarrow{e_3})\\
n &=&\frac{1}{2} (\Omega - \overrightarrow{e_3})\\
m &=&\frac{1}{2} (\overrightarrow{e_1} + i \overrightarrow{e_2})\\
\overline{m} &=&\frac{1}{2} (\overrightarrow{e_1} - i \overrightarrow{e_2})\\
\end{array}
\end{equation}

Up to the coefficients, those are also the Newman-Penrose null tetrads \cite{NewmanPenrose}.

The multiplication table for $\{l, n, m, \overline{m} \}$ is:
\begin{equation}
\begin{tabular}{|c|c|c|c|c|c|}
\hline
$~~\beta~~$ && $~~l~~$ & $~~\overline{m}~~$ & $~~m~~$ & $~~n~~$ \\
\hline
\hline
$l$ && $l$ & $0$ & $m$ & $0$ \\
\hline
$\overline{m}$ && $\overline{m}$ & $0$ & $n$ & $0$ \\
\hline
$m$ && $0$ & $l$ & $0$ & $m$ \\
\hline
$n$ && $0$ & $\overline{m}$ & $0$ & $n$ \\
\hline
\end{tabular}
\end{equation}

This multiplication table was first obtained in 1882 by Benjamin Pierce \cite{Pierceref} and was named algebra $g_4$. 

\subsection{Quantions: a mixed relativity and quantum mechanics object}

In quantum field theory an important theorem is the $CPT$ theorem. This theorem mixes quantum mechanics and relativity concepts. Complex conjugation and charge are properties of the quantum theory, and parity and time are relativity concepts. Since the quantionic algebra $\mathbb{D}$ is the only possible mathematical structure that structurally unifies relativity with quantum mechanics, the $CPT$ theorem arises naturally from it via the group of discrete transformation for quantions. 

A real quantion is defined as 
\begin{math}
p = \left(
\begin{tabular}{cc}
$r$ & $z^*$\\
$z$ & $s$\\
\end{tabular}
\right)
\end{math} where $r, s \in \mathbb{R}$ and $z \in \mathbb{C}$. Expressing $r$, $s$, and $z$ in terms of four real variables: $p_0$, $p_1$, $p_2$, $p_3$:
\begin{equation}
\begin{array}{rcl}
r &=&p_0 +  p_3\\
s &=& p_0 - p_3\\
z &=& p_1 + i p_2\\
\end{array}
\end{equation}
one has:
\begin{equation}
(p,p) = {p_0}^2 - {p_1}^2 - {p_2}^2 - {p_3}^2
\end{equation}
and
\begin{equation}
{\left(
\begin{tabular}{cc}
$r$ & $z^*$\\
$z$ & $s$\\
\end{tabular}
\right)}^{-1} = \frac{1}{(p,p)} \left(
\begin{tabular}{cc}
$s$ & $-z$\\
$-z^*$ & $r$\\
\end{tabular}
\right)
\end{equation}

Quantions are not a division algebra, and the real quantions that lack an inverse are the null rays in the Minkowski cone. Having an inverse is not a mandatory property in quantum mechanics. An easy way to see this is the fact that we do not divide by the wavefunctions directly. In the case of perturbation theory, Feynman diagrams, and propagators, one deforms the integration contour to avoid exactly the points where quantions do not have an inverse.
 
\section{Quantions: lifting a degeneracy of complex numbers}

Quantionic algebra was originally discovered in 1882, but its properties remained unexplored for a very long time until the quantal algebra research program rediscovered them using a systematic approach. However, there is another road that leads to quantions, this time completely in the realm of mathematics. For a long time, there was a mathematical bias towards division algebras, and the reason for this was an old Hurwitz theorem that states that there are only four normed division algebras: real numbers $\mathbb{R}$, complex numbers $\mathbb{C}$, quaternions $\mathbb{H}$, and octonions $\mathbb{O}$ \cite{HurwitzTheorem}. Probably the original appeal of the theorem stems from the restriction of the number of such algebras, as opposed to an infinite number of associative non-division algebras. However, as seen earlier, null space-time intervals do not have an inverse in quantionic algebra $\mathbb{D}$, and imposing the unnecessary division property eliminates relativity from $\mathbb{D}$, forcing us back at using complex numbers. 

But the complex numbers themselves have an additional property that can be regarded as a ``defect": they have a mathematical degeneracy of algebraic and geometrical concepts which if lifted will lead uniquely to the quantionic algebra. The algebraic norm of complex numbers is defined as:
\begin{equation}
A(z) = z z^*
\end{equation}

Expanding $A(z)$ in terms of the components $z = x+iy$, one has:
\begin{equation}
A(z) = x^2 +y^2
\label
{equationMetric1}
\end{equation}
Now Eq.~\ref{equationMetric1} can be understood as a metric ($M(z) = x^2 +y^2$) and this is a geometric concept. Since complex numbers were introduced for their property of algebraic closure, and since the metric is the trivial Euclidean metric in two dimensions, it takes a bit of effort to see $A(z)$ and $M(z)$ as really separate concepts. However, once the separation is made, straightforward algebraic analysis will lead uniquely to the quantionic algebra $\mathbb{D}$ as the only algebra that is able to lift this degeneracy \cite{GrginBook1} and has different algebraic and geometric norms. The two norms of quantions also have a remarkable physics interpretation. The algebraic property of quantions is related to standard quantum mechanics, and the geometric property is related to relativity. 

In quantionic algebra one can introduce complex conjugation $(*)$ and metric dual $(\sharp)$ as follows:
\begin{eqnarray}
q^*= \{a^*, c^*, b^*, d^* \}\\
q^{\sharp} = \{d, -b, -c, a\}
\end{eqnarray}	
where $q = \{a,b,c,d \}$.

The quantionic algebraic norm $A(q)$ is defined using standard Hermitian conjugation:
\begin{equation}
A(q) =q^* q =  \{ a^* a + b^* b, c^* a + d^* b, a^* c + b^* d, c^* c + d^* d \}
\end{equation}
and the quantionic metric norm $M(q)$ is the determinant of the quantionic matrix:
\begin{equation}
M(q) = ad - bc
\end{equation}
The inverse of a quantion is:
\begin{equation}
q^{-1} = \frac{q^\sharp}{M(q)}
\end{equation}
Since $M(q)$ may be zero, quantions are not a division algebra.

Not only $A(q) \ne M(q)$ in general, but as functions they reduce an eight-dimensional quantion to a four, and a two dimensional object respectively.
$M(q)$ is obviously a complex number and $A(q)$ is a real quantion because:
\begin{equation}
{ (A(q) ) }^*= {(q^*  q)}^* = q^* q^{**} = q^* q = A(q)
\end{equation}

$M(q)$ maps quantions to complex numbers and non-relativistic quantum mechanics, while $A(q)$  maps quantions into Minkowski four vectors, thus extracting relativity. 

By removing the algebraic-geometric degeneracy of complex numbers, quantions are the next number system in the sequence: natural numbers, real numbers, and complex numbers. Quantionic physics does not deform the Hilbert space; it only replaces complex numbers with a new number system. The unnecessary division property of complex numbers was the main hindrance in uncovering the relativity structure. Due to their uniqueness, quantions are nature's number system where a lot of physics will follow straight as mathematical theorems with no external ad-hoc justification. Another reason of calling quantions a number system is the existence of a hyperquantionic sequence. For real numbers, the Cayley-Dickson construction combines two real numbers into a complex number, four real numbers into a quaternion number, eight real numbers into an octonion number, and so forth using the powers of two. In the hyperquantionic sequence one starts with complex numbers and constructs groups of complex numbers using the powers of four. 

\section{Born and Zovko interpretation of the wave function}
Standard quantum mechanics based on complex numbers consists of several parts. First, we have the Hilbert space. Then, we need to postulate space and time as concepts outside Hilbert space. Finally, we need to add Born's interpretation of the wave function and the Schr\"{o}dinger equation. Generalizations of quantum mechanics were attempted to solve the unification problem. One approach is to uncover first the geometrical formulation of quantum mechanics \cite{AshtekarSchilling}. Hilbert space is understood as a K\"{a}hler space endowed with a symplectic and a metric structure. The starting point is the Hermitian inner product decomposition into real and imaginary parts:
\begin{equation}
<\Phi, \Psi> = \frac{1}{2 \hbar}G(\Phi, \Psi) + \frac{i}{2 \hbar}\Omega (\Phi, \Psi)
\end{equation}
with $G(\Phi, \Psi) = \Omega (\Phi, J \Psi)$, $J = G^{-1} \Omega$, and $J^2 = -1$. The space of physical states is the projective Hilbert space $CP(n) = U(n+1)/U(n) \times U(1)$ and the Schr\"{o}dinger equation describes a Killing Hamiltonian flow along $CP(n)$.

A complex number $z = x + i y$ can be represented as $z = x G + y \Omega$ where \begin{math}
G = \left(
\begin{tabular}{cc}
1 & 0\\
0 & 1\\
\end{tabular}
\right)
\end{math}
and \begin{math}
\Omega = \left(
\begin{tabular}{cc}
0 & 1\\
-1 & 0\\
\end{tabular}
\right)
\end{math}. We can see that from Born's interpretation, complex numbers occurs naturally in quantum mechanics but the interpretation of $G$ and $\Omega$  have completely different meaning when compared with the complex numbers introduced as a consequence of the composability principle. This geometric approach stems from the usual quantization procedure of replacing the Poisson brackets with commutators. What this does is to augment a symplectic structure with a metric structure resulting into a K\"{a}hler space. Born's interpretation of the function $\rho = \psi^* \psi$ as a probability density implies a positive norm which in turn guarantees a division algebra. Since quantions are not a division algebra, if one is to find deformations of quantum mechanics to obtain (structural) unification with relativity, the staring point must be the replacement of Born's interpretation with something else. In 2002, Nikola Zovko proposed a generalization of Born's interpretation. In quantionic algebra, Zovko's interpretation uses a current probability density $j = q^{\dagger} q$ with $j$ being a future oriented time-like Minkowski vector. Combining quantions with Zovko's interpretation leads to Dirac and Schr\"{o}dinger equations. Moreover, the Minkowski metric is fully contained within quantions and does not need to be postulated as an outside component.

So far we have discussed the main algebraic properties of quantions. As a $2 \times 2$ matrix, quantions has only the symmetries of the Lorenz group. To have equations of motions, we need to introduce additional degrees of freedom and the new structure requires the Riemannian space. Only in the flat case, derivations generate the Abelian group of translations and therefore the Poincar\'{e} group. The unique way to generalize quantions is using a sub-algebra of the $4 \times 4$ complex matrices in the following block diagonal form \cite{GrginBook2}:
\begin{equation}
Q = \left(
\begin{tabular}{cc}
A & 0\\
0 & A\\
\end{tabular}
\right)
\end{equation}
where
\begin{equation}
A = \left(
\begin{tabular}{cc}
z & v\\
u & w\\
\end{tabular}
\right)
\end{equation}
is a regular $2 \times 2$ quantion. This representation appears naturally from the complex number degeneracy elimination problem.

Up to a similarity transformation, $Q$ are unique generalizations of the $2 \times 2$ quantions. The extension, called the left algebra of quantions, allows derivation, limited analyticity properties, quantion-spinor complementarity, and Dirac equation \cite{GrginBook2}. The algebra of matrices $Q$ is a representation in terms of matrices the quantionic algebra, acts on ket column vectors, and has the $SU(2) \times U(1)$ electroweak symmetry. Associated with the left representation is a right representation which acts on bra row vectors and the left and right representations commute. The commutation property is equivalent with the associative property of quantions.

Those advanced topics are outside the scope of this introductory paper and interested readers should consult the {\it Structural Unification of Quantum Mechanics and Relativity} book by Emile Grgin\cite{GrginBook2}.

\section{Discussions and open problems}
The author believes that structural unification of relativity and quantum mechanics is a major milestone in understanding nature because it holds the potential to support a new physics paradigm centered on the old question of physics axiomatization. Although quantionic physics is still in the early development stages with many critical questions not yet researched, quantionic physics may put the phenomenological postulates of the Standard Model on a solid axiomatic foundation. While Emile Grgin refrains from speculations about the future and prefers to follow the math wherever it may lead, in this section the author is free to use the glimpses and insights learned from this new research area to provide discussions, conjectures and speculations. As such, math rigor will be replaced mostly by heuristic and philosophical arguments. 

 One of the major successes of quantionic physics is the fact that structural unification is only possible for a four dimensional space time obeying the Minkowski metric. Without a complete unification theory, the proof of the space-time dimensionality is incomplete, but quantionic research is a big step forward. (Outside unification approaches, the four dimensionality is singled out as the only case where Yang-Mills theories are renormalizable. Also, from the geometrical point of view, one can construct uncountably many inequivalent differential structures and have an interplay between Hodge duality and two-forms \cite{IngemarBengtsson}.) However, quantionic research is just beginning and there are many open problems. 

In quantionic physics, the natural symmetry is $U_q (1) = U(1) \times SU(2)$, and determining the origin of the strong force $SU(3)$ symmetry is an open problem currently under vigorous research. Increasing the available degrees of freedom by considering $U_q (2)$ can lead to $SU(3)$, but the question becomes why stop here and not consider for example $U_q (17)$, or any arbitrarily high number. What is the distinguishing property of $SU(3)$  from quantionic perspective? Preliminary results appear to answer fully this question, but it is premature to present them here.

In terms of quantum gravity, there are links between the $SO(2,4)$ group and loop quantum gravity \cite{Kerrick1},\cite{Kerrick2} and between twistors and string theory \cite{stringsTwistors}. The major problems of general relativity such as renormalizability, singularities, and global structure do not yet get much clarification from quantionic physics. 

Standard Model has the $U(1) \times SU(2) \times SU(3)$ symmetry, and Geoffrey Dixon proposed using the algebra $\mathbb C \otimes \mathbb H \otimes \mathbb O$\cite{GDixon}. From quantionic algebra, we can see that using only norm division algebras is not enough to construct the correct axiomatization of the Standard Model. 

\subsection{Axiomatization of physics}
After the Galilean revolution, physics became an experimental science. Now, with quantionic advances in unifying quantum mechanics and relativity, here is a bold speculation: what if nature enjoys uniqueness in the sense that four dimensional space time, general relativity, and quantum mechanics are mandatory consequences of a hypothetical theory of everything? What if all physics can be derived mathematically without the need for experiments in a post Galilean era? Since G\"{o}del's famous incompleteness theorem\cite{GodelTh}, we know that mathematics is infinite. But how about physics? Is physics axiomatizable? This is not a new question. It was first proposed in 1900 by David Hilbert as problem six of his famous twenty three problems that should define the next century of mathematics \cite{Hilbert1}. If problem six is solvable, uniqueness results are critical. 

When considering this problem, one should consider axioms that are not mere technical postulates, like for example the definition of Hilbert space, but principles that will separate the Platonic world of abstract mathematics from the real physical world. One such postulate is the composability principle discussed above. 
	
Dimensional analysis of Lie groups is a very powerful tool to prove uniqueness, and an important result was obtained in this way. In general relativity, if we demand that one needs to support local mathematical structures of infinite complexity (in other words a general ontology), then one necessarily obtains the orthogonal groups \cite{Rau1}. For ontology to be possible, orthogonal groups are required. 

Let us continue the discussion by proposing two other principles: the deformability principle and the universal truth property principle. 

The ``deformability'' principle was introduced in \cite{Rau1}. Deformability meant that the local physical structure was allowed to vary freely which corresponded to the requirement that arbitrary matter distributions should be allowed. Expressed in terms free of general relativity concepts, this principle demands the support of local mathematical structures of infinite complexity which in turn imply the existence of orthogonal groups of arbitrary signature $SO(p,q)$. The existence of time, or the transition from $SO(p,q)$ to $SO(1, n-1)$ requires yet another principle: the universal truth property.

In general in mathematics, the truth value of a statement depends on the context. For example, the statement that two parallel lines never meet is true in Euclidean geometry, and false on Riemannian geometry. The mathematical meaning of truth is coded by the Tarski theorem \cite{Tarski1} which roughly states that inside an axiomatic system, one cannot define the truth value of its own predicates. Thus, in mathematics, truth means that something is derived from axioms, while in the physical world truth is usually defined as something corresponding to reality and has a ubiquitous non-trivial (but easily overlooked) universal property. In physics, events occurring on the four dimensional event manifold are true for all observers and across all contexts. This is a remarkable property that can be shown to lead to the necessity of time as the only way to avoid self-referencing paradoxes via the Liar's paradox \cite{Florin1}.

The incompleteness theorem shows that mathematics in infinite in the sense that, at least in some cases, one can always find a new statement (or in the mathematical terminology a predicate) $p$, which cannot be proved or disproved within the existing axiomatic system. If the predicate is then added as a new axiom, the process can be repeated again in the extended axiomatic system. Since the new axiom can be added as either ($p$) or (not $p$), the process generates two new incompatible axiomatic systems. This process shows that the outside space and time Platonic world of mathematics is not only infinite, but also filled with contradictory axiomatic systems that cannot be organized into a coherent system. 

In the physical world however (which share at least the same complexity as the mathematical world since we can discover the mathematical axioms), the universal truth property (or equivalently global consistency) leads to a constraint which manifests itself as global hyperbolicity, or time.

\section{Acknowledgments} 
I would like to thank Emile Grgin for countless enlightening, stimulating, and enjoyable discussions. I would also like to thank Hrvoje Nikolic for introducing me to the quantionic research results and for suggesting to write this paper in the first place.

\end{document}